\documentstyle[aps,twocolumn,epsf]{revtex}

\begin{document}
\wideabs{\
\title{{\bf Phase separation and the possibility of orbital liquid states in CMR
manganites. A }$^{{\bf 139}}${\bf La NMR study.}}
\author{M. Belesi, G. Papavassiliou, M. Fardis, and G. Kallias}
\address{Institute of Materials Science, National Center for Scientific\\
Research ''Demokritos'', 153 10 Athens, Greece}
\author{C. Dimitropoulos}
\address{Institut de Physique Exprimentale, EPFL-PH-Ecublens, 1015-Lausanne,\\
Switzerland}
\date{\today }
\maketitle

\begin{abstract}
$^{139}$La NMR spin-lattice relaxation rate $1/T_1$ and rf enhancement
experiments provide evidence that the low temperature regime of the
ferromagnetic (FM) phase of La$_{1-x}$Ca$_x$MnO$_3$ segregates into
highly-conductive and poorly-conductive FM regions, associated with
differences in the orbital structure. Remarkably, phase separation is
accompanied with the appearance of an extra NMR signal from FM regions with
vanishingly small magnetic anisotropy. This feature has been attributed to
the appearance of regions with strong orbital fluctuations, resembling
droplets of an orbital liquid within the inhomogeneous FM matrix.
\end{abstract}


}

The origin of the electronic properties of mixed valence manganites with the
general formula R$_{1-x}$D$_x$MnO$_3$ (R=rare earth, D=Ca, Ba, Sr), poses
one of the most exciting open problems in the physics of strongly correlated
electron systems. Initially, these properties were determined in the
framework of the double exchange (DE) model \cite{Zener51}, which is based
on the strong Hund's coupling between hopping $e_g$ and underlying $t_{2g}$
electrons in successive Mn$^{4+,3+}$ sites. However, in the last years
experiments have shown that DE is inadequate for the full description of the
complex properties of these systems. For example, in the low dopping regime
an important question concerns the origin of the ferromagnetic insulating
(FMI) to ferromagnetic conductive (FMC) phase transition, which at first
sight appears to contradict the DE model. Recent $x$-ray resonant scattering 
\cite{Endoh99} and NMR \cite{Papavassiliou00} experiments have shown that
this transition is controled rather by orbital than by spin degrees of
freedom. Another question concerns the nature of the mixed FMI-FMC phase,
which has been observed close to the corresponding phase boundary \cite
{Papavassiliou00}. According to theoretical modells including orbital
ordering (OO) and Jahn-Teller distortions, such phase coexistence might
indicate the presence of electronic phase separation \cite{Yunoki98a}, \cite
{Okamoto}. However, it is still experimentally unclear whether the observed
mixed phases possess the same intrinsic electronic properties as the
constituting pure phases, or whether they form microscopically a ''third''
novel electronic state.

In this letter, we demonstrate experimentally for the first time that phase
separation, occuring in the low temperature regime of FM La based
manganites, is accompanied with a significant change of the $e_g$ electron
mobility at a local level. Specifically, we show that by cooling the high
temperature ''homogeneous'' FMC phase breaks into FMC and FMI regions with
enhanced - respectively reduced - electron mobility in respect to the high
temperature phase. Most important, the formation of the mixed phase is
accompanied with the detection of FM regions with extremely low magnetic
anisotropy, which is explained as showing the formation of regions with
strong orbital fluctuations.

In order to label the various magnetic structures according to their
electronic transport properties and OO we have employed $^{139}$La spin
lattice relaxation, and $^{139}$La radiofrequency (rf) enhancement
experiments. Particularly, we have shown that the abrupt decrease of the $%
^{139}$La NMR spin-lattice relaxation rate ($1/T_1$) in La$_{1-x}$Ca$_x$MnO$%
_3$ by decreasing temperature or increasing $x$, provides a direct measure
of delocalization of the $e_g$ electrons. On the other hand, rf enhancement
experiments may be used in order to monitor subttle changes of the magnetic
anisotropy at a local level. In such experiments the NMR signal intensity $I$
is recorded as a function of the indensity $H_1$ or the duration $\tau $ of
the applied rf field \cite{Weisman}. The obtained curves follow an
assymetric bell-shaped law with maximum at $n\gamma H_1\tau =2\pi /3$, which
allows the calculation of the rf enhancement factor $n$ \cite
{Papavassiliou97}. Considering that the rf enhancement reflects the coherent
response of the electron magnetic moments to an external rf fileld, it is
clear that any variation of the magnetic anisotropy field $H_A$ will be
directly reflected on $n$ \cite{Comment1}.

Experiments were performed on polycrystalline samples prepared by annealing
stoichiometric amounts of the corresponding oxides in air at $1300$ to $1400$
$^0C$. All samples were then characterized structurally at room temperature
with a D$500$ Siemens $x$-ray diffractometer, and magnetically with a SQUID
magnetometer. The obtained crystallographic and magnetic data were found to
be in accordance with literature. $^{139}$La NMR spectra in zero external
magnetic field were acquired by applying a two pulse spin-echo technique,
with pulse widths $t_{p1}=t_{p2}=0.6$ $\mu sec$ at very low rf power level,
due to the very strong rf enhancement that characterizes FM materials \cite
{Papavassiliou97}. The obtained $^{139}$La spectra consist of broad lines
located at $\simeq 20$ MHz, which unambiguously are attributed to FM regions
with fully polarized Mn spin octants \cite{Papavassiliou99}, \cite{Allodi98a}%
, \cite{Comment2}. $T_1$ was then measured at the peak of the spectra, by
applying a saturation recovery technique and fit with a multiexponential
recovery law as in previous works \cite{Furukawa99}.

\begin{figure}[tbp] \centering
\epsfxsize=8.6cm \leavevmode
\epsfclipon
\epsffile{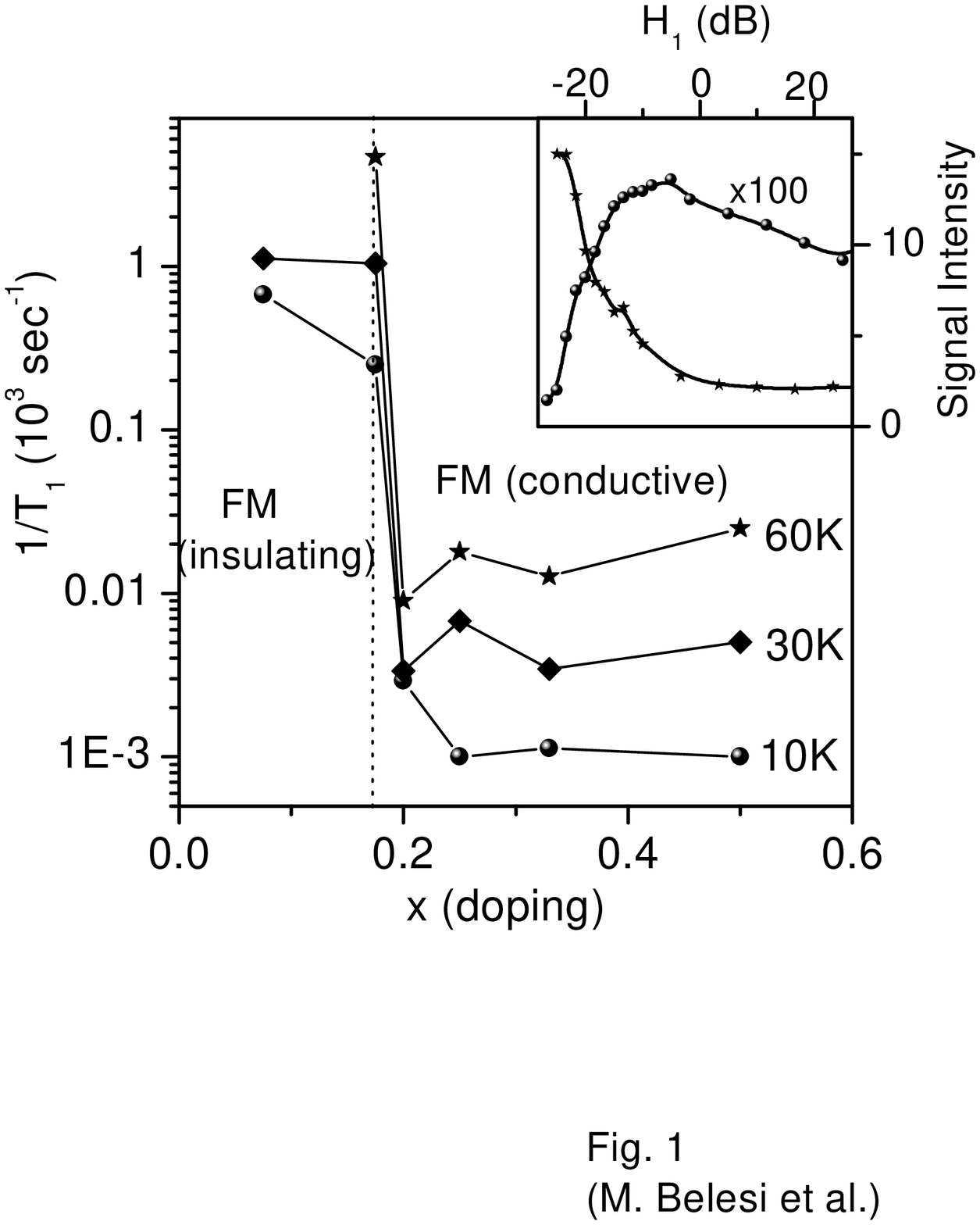}
\caption{$^{139}$La $1/T_1$ of La$_{1-x}$Ca$_{x}$MnO$_3$, as a function of $x$. The inset demonstrates
the $^{139}$La NMR signal intensity as a function of the rf field $H_1$ for $x=0.1$ ($\bullet$) and $0.33$ ($\star$). 
All spin lattice relaxation experiments were performed at $H_1=-10$dB $\simeq 3$Gauss.}
\label{fig1}%
\end{figure}

Figure \ref{fig1}, exhibits $^{139}$La $1/T_1$ vs. $x$ measurements of La$%
_{1-x}$Ca$_x$MnO$_3$ at three different temperatures. It is observed that by
increasing $x$, $1/T_1$ decreases abruptly by almost two orders of
magnitude, at $x\approx 0.175$, which according to the literature defines
the phase boundary between the FMI and FMC regimes of the $T-x$ phase
diagram. This is a remarkable result, which shows that $e_g$ electrons play
a major role in $^{139}$La spin-lattice relaxation phenomena. In order to
describe qualitatively this behaviour, we have used the general approach
invented by Warren for relaxation studies of metal-insulator transitions in
liquid semiconductors \cite{Warren71}. In general, nuclear spin-lattice
relaxation in CMR manganites is determined by fluctuations of the local
magnetic field $H_{loc}$ at the nuclear sites, (i) due to $e_g$ electron
hopping, and (ii) due to interaction with spin waves. However, in systems
with spin-polarized conduction bands, like CMR manganites \cite{Wei97}, spin
wave scattering is determined by weak two-magnon processes, and $^{139}$La $%
1/T_1$ should be dominated by direct $e_g$ electron spin flips, if
energetically allowed. The latter assumption is supported by recent tight
binding calculations in FM CMR manganites \cite{Papaconstantopoulos98},
which show that in the presence of electron-phonon coupling the minority $%
t_{2g}(\downarrow )$ band attains a small finite DOS at $E_F$, and spin flip
between $e_g(\uparrow )$ and $t_{2g}(\downarrow )$ energy states is in
principle possible. Following Warren's approximation, and by assuming an
exponentially decaying autocorrelation function for the fluctuations of $%
H_{loc}$, we may easily arrive at the expression \cite{Warren71} $%
1/T_1\simeq TN(E_F)^2\tau _e$/$\left\{ \hbar N(E_F)\right\} $, where $\tau
_e $ has the significance of the lifetime residence of $e_g$ electrons on
the Mn sites.

In case of delocalised electron states, where the electron mean-free path $%
\ell >>a$ ($a$ is the distance to the nearest neighbour), $\tau
_e=a/v_F=\hbar N(E_F)$ \cite{Warren71}, and the above relation transforms to
the standard Korringa relation, $1/T_1\propto TN^2(E_F)$. However, in case
of charge localization, i.e. for $\tau _e>>\hbar N(E_F)$, $1/T_1$ is
sufficiently higher than for delocalised electron states. Evidently, the
abrupt reduction of $1/T_1$ in Figure \ref{fig1} provides a direct evidence
of conductivity enhancement, due to delocalization of the $e_g$ electron
states. Such a behaviour is reminisence of a sharp metal-insulator
transition of the Anderson type, and not of a percolative transition as
proposed by many authors. Most notably, the onset of metallicity for $x\geq
0.175$ is characterized by a strong amplification of the rf enhancement
factor, as recently reported by Dho et al. \cite{Dho99a}. This is clearly
observed in the inset of Figure \ref{fig1}, which exhibits $^{139}$La NMR $I$
vs. $H_1$ curves of La$_{1-x}$Ca$_x$MnO$_3$ for $x=0.1$, and $0.33$, at $T=5$%
K. For $x=0.33$ the maximum of the $I$ vs. $H_1$ curve shifts to extremely
low $H_1$ values, which indicates a considerable reduction of the magnetic
anisotropy \cite{Comment1}. Considering that (i) $H_A$ is mainly determined
by the $LS$ coupling, and (ii) according to previous works on Sr-dopped
systems \cite{Endoh99} the metal-insulator transition correlates with OO
changes, we argue that the observed change in the rf enhancement reflects a
substantial modification in the orbital structure.

\begin{figure}[tbp] \centering
\epsfxsize=8.6cm \leavevmode
\epsfclipon
\epsffile{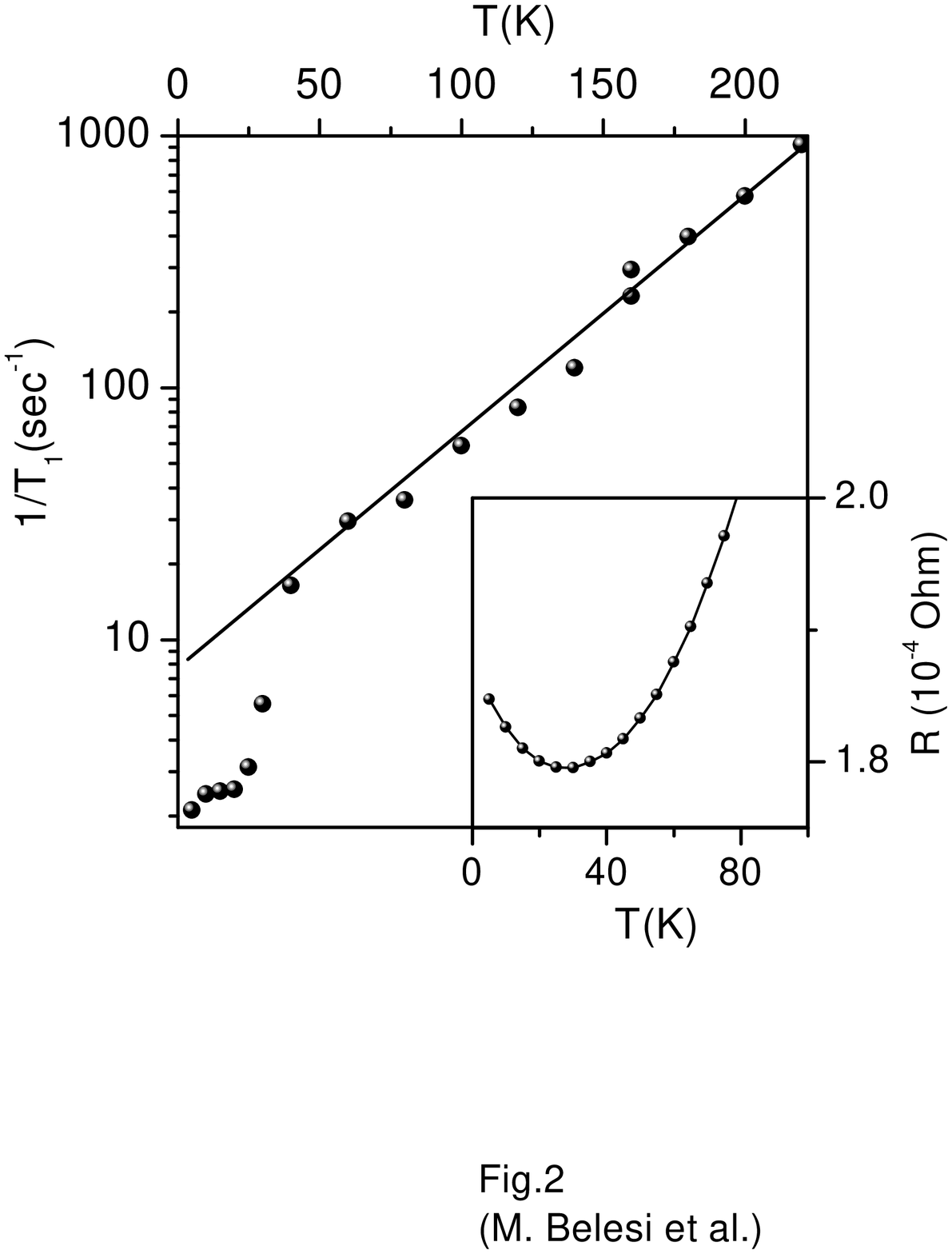}
\caption{ $^{139}$La $1/T_1$ as a function of temperature of
La$_{0.75}$Ca$_{0.25}$MnO$_3$. The inset exhibits the corresponding $R$ vs.
$T$ curve.}\label{fig2}%
\end{figure}

Figure \ref{fig2} demonstrates $^{139}$La $1/T_1$ vs. $T$ measurements for $%
x=0.25$. It is observed that by decreasing temperature $1/T_1$ decreases
almost exponentially, in agreement with previous works in relevant materials 
\cite{Savosta99}. For $T<30$K $1/T_1$ exhibits an almost stepwise decrease,
which may be assigned to enhancement of the $e_g$ electron hopping below
that temperature. Remarkably, the steep descent of $1/T_1$ is accompanied
with the appearance of $^{55}$Mn NMR signal from FMI regions, as shown in
ref. \cite{Papavassiliou00}. Such FMI regions can not be detected by $^{139} 
$La NMR for $x\geq 0.175$, because signals from FMI regions with extremely
low $T_1$s are masked from the much stronger signals from FMC regions.
Apparently, the overall resistivity enhancement below $30$K (inset of Figure 
\ref{fig2}) may be attributed to restriction of the conductivity channels
below that temperature. These results demonstrate that for $T\leq 30$K the
system splits into regions with enhanced - respectively reduced -
conductivity in comparison to the high temperature FMC phase.

\begin{figure}[tbp] \centering
\epsfxsize=8.6cm \leavevmode
\epsfclipon
\epsffile{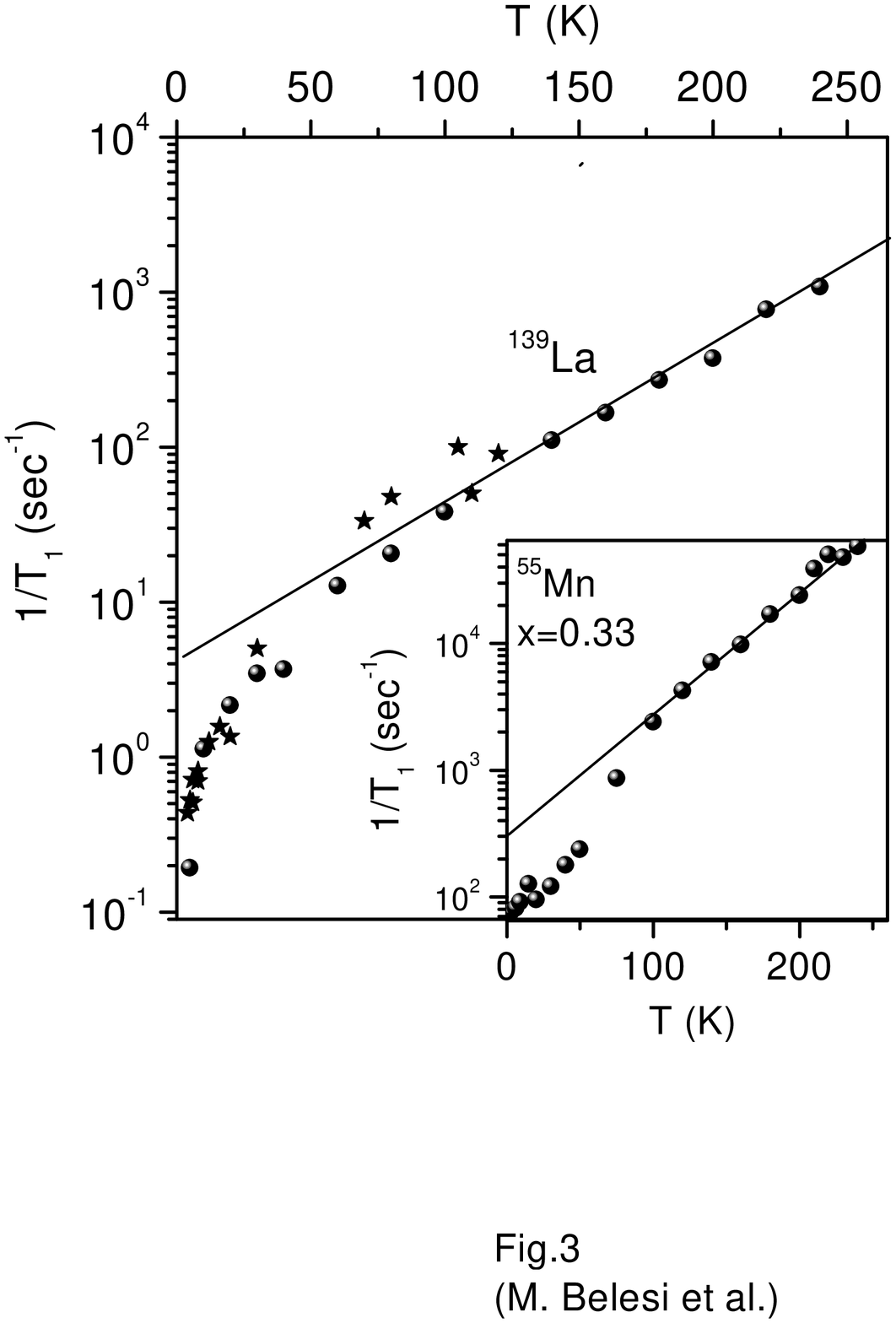}
\caption{ $^{139}$La $1/T_1$ vs. T for La$_{0.67}$Ca$_{0.33}$MnO$_3$ ($\bullet$) and
La$_{0.5}$Ca$_{0.5}$MnO$_3$ ($\star$). The inset exhibits $^{55}$Mn $1/T_1$ for
$x=0.33$.} \label{fig3}%
\end{figure}

A similar $1/T_1$ behaviour is followed by $x=0.33$ and $0.5$, as observed
in Figure \ref{fig3}. For reasons of comparison, the $^{55}$Mn $1/T_1$ vs. $%
T $ curve for $x=0.33$, measured at the peak of the FMC signal at $\nu
\simeq 375$ MHz \cite{Papavassiliou00}, is also shown in the inset of Figure 
\ref{fig3}. In case of $x=0.33$, the deviation from the high temperature $%
1/T_1$ vs. $T$ law is realized below $80$K, whereas for $x=0.5$ a similar
behaviour is observed; however, the experimental error of the $1/T_1$ data
points above $100$K does not allow precise determination of the pertinent
temperature, which defines the inflection point. It is worth to notice that
for $x=0.5$ the reduction of $1/T_1$ at low temperatures takes place deep
into the antiferromagnetic phase, hence it concerns enhancement of the
electron mobility within the FMC islands that persist below $T_N$ \cite
{Papavassiliou99}, \cite{Allodi98a}.

\begin{figure}[tbp] \centering
\epsfxsize=8.6cm \leavevmode
\epsfclipon
\epsffile{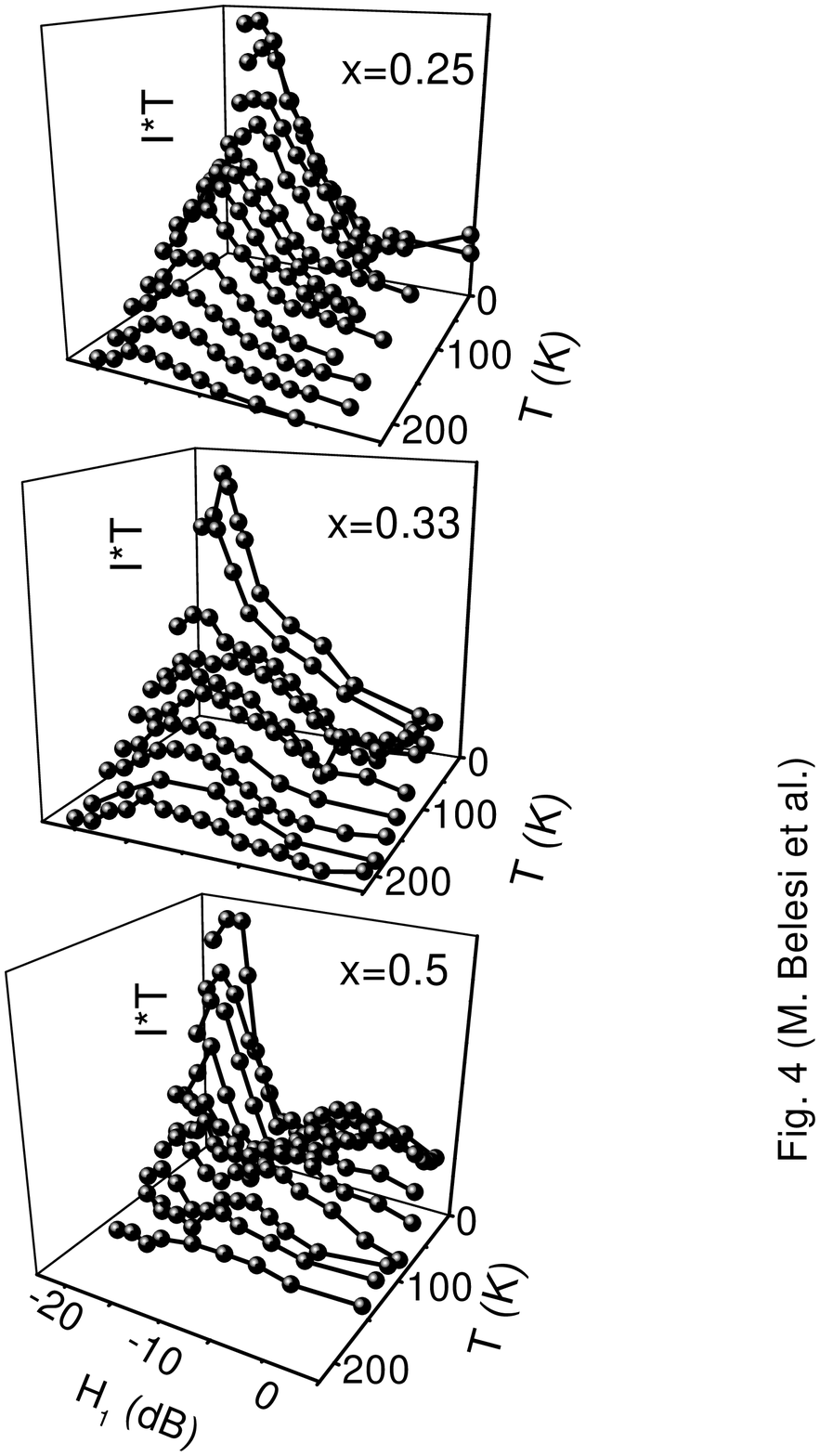}
\caption{$^{139}$La NMR signal intensity as a function of the rf field $H_1$ at various temperatures, 
for $x=0.25$, $0.33$, and $0.5$.} \label{fig4}%
\end{figure}

In order to examine whether the temperature dependence of the local electron
mobility is associated with modifications of the OO, detailed $I$ vs. $H_1$
measurements for $x=0.25$, $0.33$, and $0.5$ were performed as a function of
temperature (Figure \ref{fig4}). In case of $x=0.25$ and $0.33$, by
decreasing temperature the maximum of the $I$ vs. $T$ curves is clearly
observed to shift to lower $H_1$ fields. Remarkably, the abrupt reduction of 
$1/T_1$ is accompanied with the appearance of a second maximum in the $I$
vs. $H_1$ curves at extremely low $H_1$ values. This is a direct evidence
that the low temperature regime of the FMC phase contains at least two
distinct FM electron spin configurations, differing in $H_A$ and therefore $%
OO$ \cite{Comment3}. By increasing $x$, the low-$H_1$ maximum remains
unshifted, whereas the initial maximum shifts to higher rf power levels. It
is also notable that for $x=0.25$ and $0.5$, the second maximum appears
simultaneously with the onset of the mixed phase, i.e. below $\simeq 30$K
for $x=0.25$ \cite{Papavassiliou00}, and $\simeq 140$K for $x=0.5$ \cite
{Allodi98a}. These results corroborate with (i) recent $^{119}$Sn Moessbauer
measurements in $1\%$ Sn-doped La$_{0.5}$Ca$_{0.5}$MnO$_3$, which indicate
the appearance of a third FM signal component for $T<120K$ \cite
{Simopoulos00}, and (ii) magnetoresistance experiments in thin films of La$%
_{0.67}$Ca$_{0.33}$MnO$_3$ \cite{Donnell00}, which show that below $100$K
magnetoresistance anisotropy changes rapidly from two-fold to four-fold.

In view of the up-to-now discussion, the following picture may be envisaged:
For $x\geq 0.175$ in the low temperature regime of the $T-x$ magnetic phase
diagram the system breaks into FMC regions with high conductivity and FMI
regions with low conductivity, which differ mainly in the orbital structure.
In the FMC regions more than one orbital configurations may be realized, as
implied by the subsequent shift of the prime maximum in the $I$ vs. $H_1$
curves, by varying dopping or temperature. This is in conformity with recent
theoretical calculations \cite{Maezono98}, which have shown that by
increasing $x$ the FM state is realized, whereas OO may change continuously
from orbital $G$ $(x^2-y^2)/(3z^2-r^2)$ near $x=0$, to orbital $C$ $%
(x^2-y^2)/(3z^2-r^2)$ for $x\approx 0.3$, and orbital A $%
[(3z^2-r^2)+(x^2-y^2)]/[(3z^2-r^2)-(x^2-y^2)]$ for $0.4\leq x\leq 0.7$.
Remarkably, the calculated energy difference among these states is very
small \cite{Maezono98}, suggesting the presence of strong orbital
fluctuations. In such a case, the formation of islands of an orbital liquid
with extremely low $H_A$ is possible. This might give an explanation to the
appearance of the maximum at extremely low $H_1$ field in the $I$ vs. $H_1$
experiments. In this scenario, by increasing temperature the onset of
Jahn-Teller distortions \cite{Louca97} might lift the degeneracy of the
various states, and the orbital liquid would freeze in the high temperature
FMC orbital configuration.

In conclusion, $^{139}$La $1/T_1$ and rf enhancement measurements in
La-based manganese perovskites provide clear evidence that differences in OO
underly phase segregation tendencies into FM states with high (respectively
poor) $e_g$ electron mobility. At low temperatures where the effect of
Jahn-Teller distortions is minimized, NMR experiments suggest that
degeneracy in the energy of these states may lead to the appearance of
orbital liquid regions, in agreement with recent theoretical predictions 
\cite{Maezono98}.

{\it Acknowledgment }This work has been partially supported by the INTAS
project 97-30253.


\begin{references}
\bibitem{Comment1}  In FM domains $n$ is related to $H_A$ according to
formula \cite{Weisman} $n=H_{loc}/(H_{ext}+H_A)$, where $H_{loc}$ is the
local magnetic field at the position of the La or Mn nuclei and $H_{ext}$
the externally applied magnetic field. Similarly, in FM domain walls by
increasing $H_A$ the rf enhancement is expected to reduce. Hence,
contribution from regions with different magnetic anisotropy will produce
multiple maxima in the $I$ vs. $H_1$ curve.
\bibitem{Comment2}  In case of $x=0.5$, $^{139}$La NMR signals from AFM spin
configurations in zero external magnetic field, if observable, should be
located at very low frequencies, due to cancellation of the transferred
hyperfine field at the position of the La nuclei \cite{Papavassiliou99}.
\end{references}
\end{document}